| | |
|---|---|
| **Abstract** | Two atmospheric pressure plasma jet devices – a plasma gun and a plasma Tesla jet – are compared in terms of safety and therapeutic efficiency to reduce the tumor volume progression of cholangiocarcinoma, i.e. a rare and very aggressive cancer emerging in biliary tree. For this, a three steps methodology is carried out. First, the two APPJ have been benchmarked in regard to their electrical and physico-chemical properties while interacting with material targets: dielectric plate, liquid sample, metal plate and an equivalent electrical circuit of human body. The propagation properties of the ionization wave interacting with these targets are discussed, in particular the profile of the related pulsed atmospheric plasma streams. In a second step, a dermal toxicity survey is performed so as to define an experimental operating window where plasma parameters can be changed without damaging healthy skin of mice during their exposure to plasma and without inducing any electrical hazards (burnings, ventricular fibrillation). Optimal conditions are identified discarding the conditions where slight alterations may be evidenced by histology (e.g. prenecrotic aspect of keratinocytes, alterations in the collagen structure). Hence, for the two APPJ plasma parameters these conditions are as follow: duty cycle=14 %, repetition frequency=30 kHz, magnitude=7 kV, gap=10 mm and exposure time=1 min. In a third step, the two plasma jets are utilized on cholangiocarcinoma xenograft tumor model developed in immunodeficient mice. The two devices are safe and a significant therapeutic efficiency is demonstrated with the plasma Tesla. In conclusion, we have developed a safe cold atmospheric plasma device with antitumoral properties in preclinical model of cholangiocarcinoma, opening the path for new anticancer treatment opportunities. |

# I. Introduction

[**DBD: a large panel of configurations**] Cold atmospheric plasmas are weakly ionized gases containing energetic and chemical transient species (electrons, ions, metastables, radicals) while presenting radiation, gas flowing and electromagnetic field properties (Brandenburg et al 2018). In laboratory, they can be easily generated by supplying electrical power to a device containing either one or two electrodes. The electrode connected to the high voltage power supply is referred as the exciting electrode while the second electrode is brought to the ground and referred as the counter-electrode. As a result a high magnitude electric field can be generated to create electrical discharges that partially ionize the gas into cold plasma. Of the most commonly cold plasma devices used in laboratories, the Dielectric Barrier Discharge (DBD) appears ubiquitous owing to its low manufacturing and implementation costs, as well as its great versatility in regard of the many diversified applications like ozone generation, incoherent excimer UV radiation, air purification, surface modifications, etc. (Kogelschatz et al 1997) (Laroussi & Akan 2007). In these DBDs, one electrical insulating layer, typically a dielectric material like quartz or alumina is utilized as a barrier to prevent arcing from plasma current. Whatever their 1 or 2 electrode(s) configurations, two types of DBD can be distinguished: (i) the non-flowing DBDs where the powered electrode is enwrapped by an insulator material and where plasma remains confined in the interelectrode region (ii) gas-flowing DBDs where plasma is located in this region as well as further away: in the post-electrode region. There, the ionized gas is referred as a plume to design its emissive properties although long lifetime radicals can propagate much further away in the post-discharge, as sketched in Figure 2. The gas-flowing DBDs are more commonly referred as plasma jets and more specifically as Atmospheric Pressure Plasma Jets (APPJ) if they operate in ambient air (Isbary et al 2013). As illustrated in Figure 1, one can distinguish two types of APPJ configurations: (i) APPJ devices with a single metal electrode biased to the exciting potential (typically high voltage) while the counter-electrode is the biological target (grounded or floating potential) exposed to plasma, (ii) APPJ devices with two metal electrodes (exciting electrode and counter electrode) while the biological target can eventually play a role of third electrode . Among the "two-electrodes APPJ" successfully applied upon *in vivo* experiments, the plasma gun (PG) is composed of an outer ring electrode as sketched in Figure 1 and an inner pin electrode directly in contact with the gas/plasma (Darny et al 2017) (Robert et al 2012) (Robert et al 2009). Another configuration of interest, referred in this article as Plasma Tesla Jet (PTJ), presents two ring electrodes located on the outer tube.

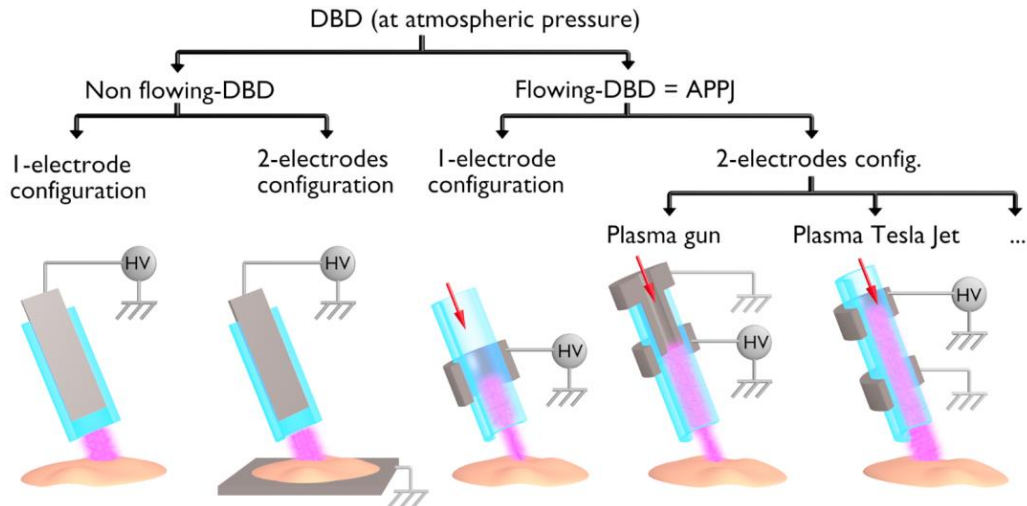

***Figure 1. Cross-section views of dielectric barrier discharge devices treating a biological tissue. For flowing-DBDs (APPJ), the red arrows indicate the direction of carrier gas flow.***

[**Target configurations**] The physico-chemical properties of a cold plasma do not solely depend on the DBD device itself but also on the biological target under exposure (e.g. tissue, tumor, skin). Understanding this device-target interaction is crucial owing to its significant effects on the plasma properties, as already shown on fluid-dynamics (Li et al 2017) (Boselli et al 2014) (Robert et al 2012), on electrical parameters (Li et al 2017), on electromagnetic field (Darny et al 2017) and on deposited electrical charge distribution (Wild et al 2013) (Stoffels et al 2008). These interactions have been extensively investigated on metal plate at ground potential (Darny et al 2017) (Li et al 2017), on dielectric substrate (Li et al 2017) (Boselli et al 2014) and on liquids (Li et al 2017).

[**APPJ applied to medicine**] From 2005 until today, these DBD devices have been utilized for medical applications and represent approximately 200 original articles (Dubuc et al 2018). 95 % of these publications deal with *in vitro* treatments performed on tumor cell lines (e.g. brain, lung, blood, cervical melanoma and breast cancer) while less than 5 % report *in vivo* experiments carried out on murine models. Besides, only 3 completed clinical trials report long term plasma-effects on human cancer although no antitumor effects have been demonstrated so far (Metelmann et al 2018) (Schuster et al 2016) (Metelmann et al 2015) (Hoffmann et al 2010). The small number of *in vivo* studies illustrates the challenge of generating a plasma safe for the patient and therapeutically efficient, in particular in the cancer field. So far, the PG appears as a good candidate owing to its success in treating pancreatic cancer (Brullé et al 2012) and melanoma (Binenbaum et al 2017) in murine models and, tissue oxygenation (Collet et al 2014). The PTJ has already been applied once on bladder tumor upon *in vivo* experiment (Keidar et al 2011) and is investigated here on another cancer, the cholangiocarcinoma.

[**Cholangiocarcinoma**] Cholangiocarcinoma is a heterogeneous group of aggressive malignancies that can emerge at every point of the biliary tree from the canals of Hering into the liver to the main bile duct. CCA is the second most frequent type of primary liver cancer and ~3% of all gastrointestinal neoplasia.

Cholangiocarcinoma are generally asymptomatic in early stages, they are diagnosed when the disease has already metastasized, drastically complicating their therapeutic treatment options (Banales et al 2016). Surgical resection is the only effective therapy, but it can only be applied in 20 % of patients and the 5-year survival rate remains as low as 15-40 %. Most of the patients who cannot benefit from surgery undergo a palliative treatment with a combination of gemcitabine and oxaliplatin platinum salt (GEMOX), the only chemotherapy validated for advanced unresectable CCA (Valle et al 2010). In case of tumor progression after this first line of treatment, there is no other treatment approved to date. Tumor size and other features (anatomical location, vascular and lymph node invasion and metastasis) condition the potential surgical and/or radiological options but chances of recurrence are very high. Owing to these limitations, the emergence of new therapeutic options is eagerly needed.

[**Overview of this article**] This article is divided into 3 stages.

- First, two APPJ devices have been engineered - a plasma gun and a plasma Tesla jet - to measure the electrical properties of plasma with/without material target. Studying the device-target interaction is a preliminary but essential step before carrying out *in vivo* experiments since it allows to easily and regularly calibrate the two APPJ devices as well as to compare their properties with plasma sources engineered by other teams. In that respect, 5 target configurations are considered in this work: APPJ in free jet and APPJ treating several types of material targets located 10 mm away.

- Second, *in vivo* experiments are carried out in mice with PG and PTJ to verify the absence of any toxic effects induced on skin.

- Third, the therapeutic efficiency induced by PG and PTJ is studied using a cholangiocarcinoma xenograft tumor model developed in immunodeficient mice.

# II. Experimental setup

## II.A. Plasma sources

Two atmospheric pressure plasma jet (APPJ) devices have been compared in regard of physico-chemical properties as well as therapeutic efficiency on murine models upon in vivo experiments: a Plasma Gun (PG) (Darny et al 2017) (Robert et al 2012) (Saron et al 2011) and an alternative configuration named Plasma Tesla Jet (PTJ). Schematic views of the PG and PTJ are depicted in Figure 2a and 2b, respectively. The two APPJ generate plasma in a 10 cm long dielectric quartz tube with a 4 mm inner diameter and a tube thickness of 2 mm.

The major difference between PG and PTJ relies on their respective electrode configurations:
- In the PG, a 50 mm long inner electrode is centered in the tube and supplied with high voltage. A 10 mm wide grounded ring electrode is set on the outer quartz tube. Along the tube axis, the middle of the ring electrode corresponds to the end of the coaxial HV electrode, as sketched in Figure 2a.
- In the PTJ, two ring electrodes, 10 mm long, are set on the outer surface of the quartz tube and separated by a distance of 10 mm. One electrode is connected to the ground while the other is biased to the high voltage.

In both APPJ, a distance of 50 mm separates the exit of the quartz tube and the down part of the exciting electrode as mentioned in Figure 2.

## II.B. Electrical environment & diagnostics

The two APPJ are supplied in helium (1000 sccm) and powered by the same mono-polar square pulse high voltage generator (Spellman, SLM 10 kV 1200 W) coupled with a Smart HV Pulses Generator (RLC electronic, NanoGen1 10 kV). To measure their electrical parameters, two capacitors are placed downstream and upstream of the APPJ as sketched in Figure 3a: (i) the upstream capacitor ($C_{m1}$) is placed between the high voltage generator and the exciting electrode to measure the total current provided by the generator, (ii) the downstream capacitor ($C_{m2}$) is placed between the counter electrode and the ground potential. Wall capacitors are represented by $C_W$ while gas capacitance is represented by $C_{IE}$ in the interelectrode region and $C_{PE}$ in the post-electrode region as shown in Figure 3a.

5 targets configurations are considered: APPJ in free jet (i.e. target infinitely remote from the device) and APPJ treating material targets located 10 mm away. Here, 4 types of targets (area 100 cm²) are studied: a metal plate at floating potential, a dielectric plate, an aqueous liquid (floating potential, conductivity 650 µS/cm) and an equivalent electrical human body (EEHB) circuit whose specifications are detailed in (Judée et al 2019).

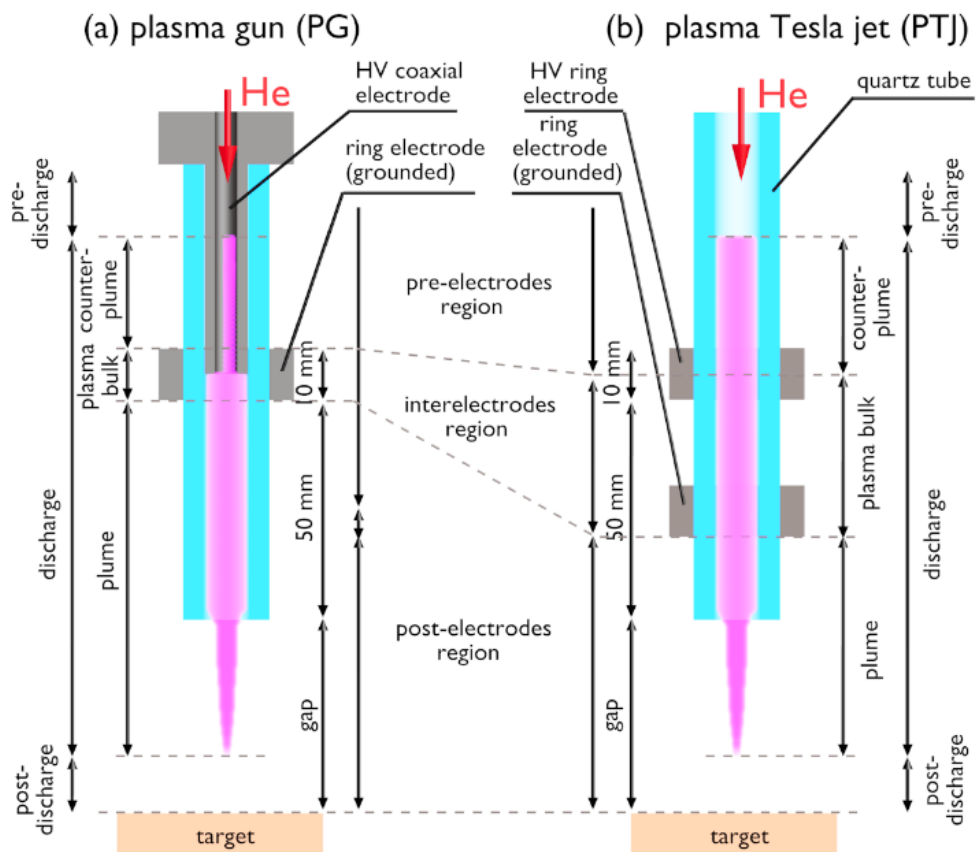

***Figure 2. Experimental setup of (a) Plasma Gun device (PG) and (b) Plasma Tesla Jet device (PTJ).***

Total power (delivered by the high voltage generator), plasma power and target power are estimated using high-voltage probes (Tektronix P6015A 1000:1, Teledyne LeCroy PPE 20 kV 1000:1, Teledyne LeCroy PP020 10:1) and an analog oscilloscope (HMO3004, Rohde & Schwarz). All currents and plasma powers are deduced according assumptions from (Judée et al 2019) and equivalent electrical circuits introduced in Figure 3.

## II.C. Equivalent electrical circuits

The equivalent electrical circuits of the plasma gun and plasma Tesla jet are introduced in Figure 3a and 3b respectively while the equivalent electric models of the 5 aforementioned configurations are represented in Figure 3c. The "no target" configuration is modeled as an APPJ interacting with an infinitely distant target. Its resulting equivalent electrical circuit is a single capacitor of 0 Farad to represent the absence of collected charges. Similarly, the "dielectric target" can be modeled by two capacitors in series: the capacitor $C_{diel}$ is specific to the material properties of the target and (non-null value) while $C_{air}$ is 0 Farad and represents the absence of charge transfer between the floating target and the ground. The same approach is followed with the "ungrounded conductive target" where a resistor is in series with the air capacitor. Here the resistance is specific to the material properties of the target. Finally, the EEHB (Equivalent Electrical Human Body) configuration corresponds to a resistor (1500 $\Omega$) in parallel with a capacitor (100 pF). It represents the electrical response of human body to electrical stimuli as detailed in (Judée et al 2019).

The electrical power transferred to the targets cannot be determined in all the configurations. With the EEHB target, it can easily be deduced from $P = f. \int_T V_{target}.I_{target}.dt$ by measuring $V_{target}$ with the HV probe and $I_{target}$ by applying Ohm's law to the grounded resistor (1500 $\Omega$). In the other configurations, target currents cannot be measured: (i) in free jet or in the dielectric target, only capacitors are included in the models sketched in Figure 3c. By definition, a capacitor can only store reactive power in electrostatic or magnetic form and cannot consume active power, hence resulting in $P_{target}$=0, (ii) in ungrounded conductive targets (metal, water), the model is composed of a resistor in series with a capacitor. The electrical power can only be dissipated as "active power" in the resistor but remains unknown since the current cannot be estimated experimentally.

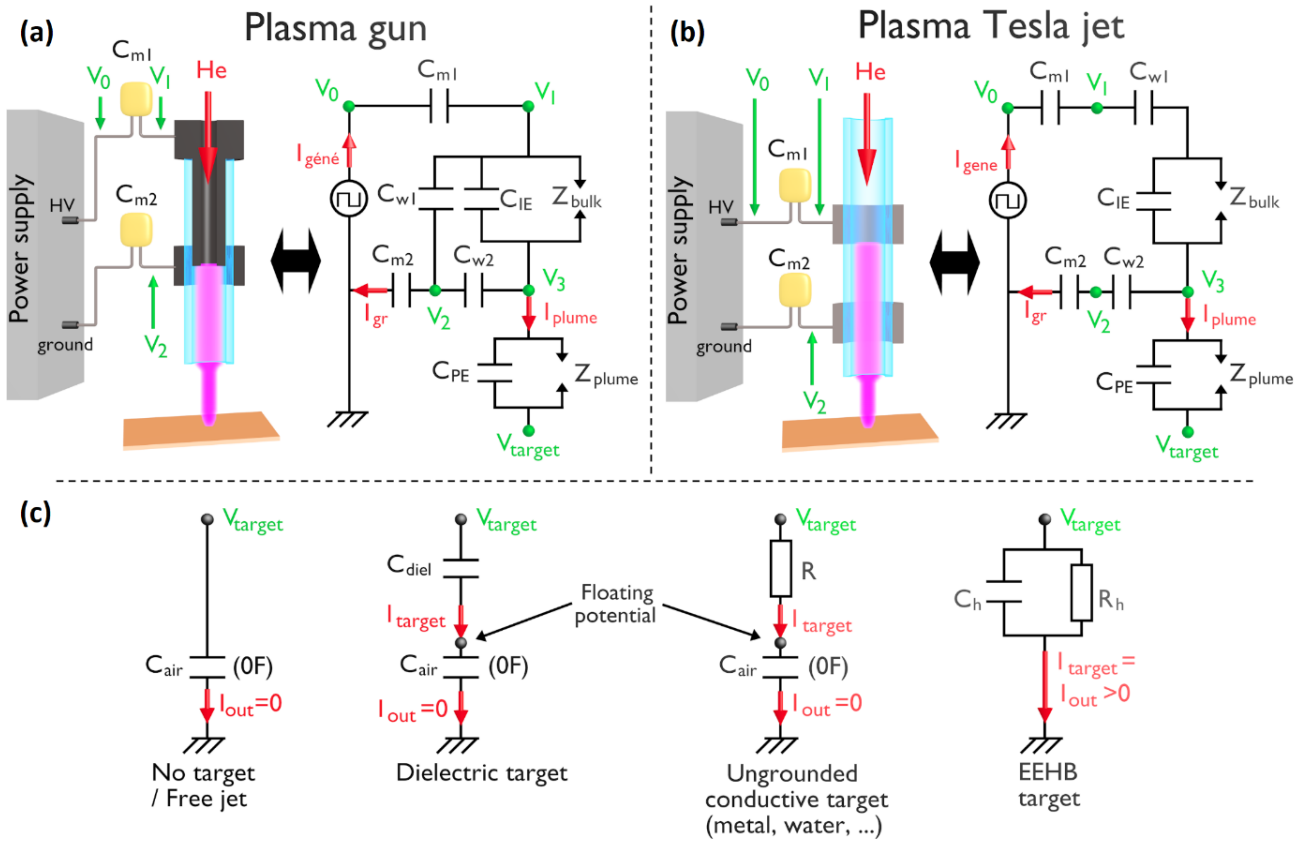

**Figure 3.** *Equivalent electrical circuits of (a) plasma gun (PG), (b) plasma Tesla jet (PTJ) and (c) targets configurations.*

## II.D. Optical emission spectroscopy

The radiative emission of the plasma jet, 300-800 nm, is collected by an optical emission spectrometer (Andor SR-750-B1-R) operating in the Czerny Turner configuration with a focal of 750 mm. It is equipped with an optical fiber (Leoni fiber optics SR-OPT-8014, 100 µm diameter) and an ICCD camera (Andor Istar, 2048 x 512 imaging array of 13.5 µm x 13.5 µm pixels). Diffraction is performed using a 1200 grooves/mm grating in the visible range. Due to low emissivity of plasma, OES spectra are acquired along the jet axis, i.e. side-on. Moreover, a converging lens (ThorLabs, LA4380-UV, f=100 mm) is placed between the plasma plume and the optical fiber to focus and collect a maximum of plasma emission (see inset of Figure 8). Finally, a high pass optical filter (Newport 10CGA-225) is placed between the optical fiber and the plume to eliminate lines and bands of the second order diffraction.

## II.E. Mass spectrometry

Gas phase analysis is completed using a quadrupole-based mass spectrometer (Model HPR-20 from Hiden Analytical Ltd.). Plasma chemical species are collected by a quartz capillary whose inlet is fixed onto a 2-axis stages plate to perform spatial profiles of the APPJ (radially and axially) with a resolution of 1 mm. This capillary is 1 m long, flexible, chemically inert and heated at 200 °C to prevent chemisorption. Then, a three-stage differentially pumped inlet system separated by aligned skimmer cones and turbo molecular pump, enables a pressure gradient from $10^5$ bar to $10^{-7}$ bar at the entrance of the ionization chamber. There, ionization energy is set at 70 eV. The residual gas analyzer (RGA) detector is used for scanning masses from 1 to 50 amu.

In all experiments, APPJs are supplied with helium gas. In the plume region, helium interacts with ambient air to form a gas mixture of pure helium, dry air and water vapour. Each of these components is characterized by its gas molar fraction χ expressed as the ratio of its partial pressure to the total pressure measured inside the ionization chamber. Dry air partial pressure is assumed to be the sum of oxygen, nitrogen, argon and carbon dioxide partial pressures. In this article, all the gas molar fractions are calculated over averaged treatment times (typically 1 min) rather than accounting on a single pulse.

## II.F. In vivo experimentation

[**Dermal toxicity test**] Innocuity of PG and PTJ is tested on the skin of anesthetized 5-weeks-old female ATHYM-Foxn1 nu/nu mice (Janvier Labs, France). Several plasma treatments are performed with various exposure times (1, 5, 10 min) and two values of duty cycles (14% and 24%). Then, the skin is harvested, fixed in 10 % formalin, embedded in paraffin and stained with hematoxylin-eosin (HE) for histological analysis (Suvarna et al 2018. Animal experiments have been performed in accordance with the French Animal Research Committee guidelines and all procedures have been approved by a local ethic committee (No 10609).

[**Cell culture**] EGI-1 cancer cells, derived from extrahepatic biliary tract, are obtained from the German Collection of Microorganisms and Cell Cultures (DSMZ, Germany). Cells have been cultured in DMEM supplemented with 1 g/L glucose, 10 mmol/L HEPES, 10 % fetal bovine serum (FBS), antibiotics (100 UI/mL penicillin and 100 mg/mL streptomycin), and antimycotic (0.25 mg/mL amphotericin B; Invitrogen). Cells have been routinely screened for the presence of mycoplasma and authenticated for polymorphic markers to prevent cross-contamination.

[**Xenograft tumor model**] Animal experiments are performed in accordance with the French Animal Research Committee guidelines and all procedures approved by a local ethic committee (No 10609). $2 \times 10^6$ of EGI-1 cells are suspended in 60 µL of PBS and 60 µL of Matrigel® growth factor reduced (Corning) and implanted subcutaneously into the flank of 5-week-old female ATHYM-Foxn1 nu/nu mice (Janvier Labs, France). Mice are housed under standard conditions in individually ventilated cages enriched with a nesting material and kept at 22 °C on a 12 h light/12 h dark cycle with ad libitum access to food and tap water. Tumor growth is monitored by measuring every 2-3 days the tumor volume ($V_{xenograft}$) with a caliper as follows: $V_{xenograft} = x.y^2/2$ where x and y are the longest and shortest lateral diameters respectively. Once tumor volume reaches approximately 200 mm³, plasma treatments are initiated: two treatments, 7 days distant, with the PG and two other treatments, 7 days distant, with the PTJ.

[**Statistical analysis**] Results were analyzed using the GraphPad Prism 5.0 statistical software. Data are shown as means standard error of the mean (SEM). For comparisons between two groups nonparametric Mann–Whitney test were used.

# III. Results

## III.A. Electrical study of plasma

Plasma current is usually measured through a capacitor (or resistor) placed between the counter electrode and the ground potential (Fang et 2016) (Kostov et al 2009). If such approach is acceptable for 2 electrode configurations, it can barely be considered where (biological) target can play a role of charge collector or of third electrode. Here, the current provided by the generator is the sum of the current measured on the ground electrode and of the current measured on the target. Besides, electrical plasma power corresponds to the difference between the power delivered by the generator (measured at $C_{m1}$) and the electrical power deposited on the target.

In Figure 4, the electrical plasma power of PG and PTJ is compared as a function of the plasma voltage, considering the 5 aforementioned configurations (free jet, metal target, dielectric target, water target and EEHB target). As shown in Figure 4a, the PG electrical power increases with $V_0$: from 1 kV to 6 kV, it increases from 0.2 W to 9.0 W for all configurations except with the EEHB target where a value close to 20 W is reached. Then, for higher values of $V_0$, e.g. 9.0 kV, the plume can create a direct electrical contact with the target. Hence, in free jet or with the dielectric target, $P_{plasma}$=18.0 W while higher values are obtained for the conductive targets, namely 26.5 W with the metal plate and 33.0 W with the water sample. For the EEHB target considering the same voltage of 9.0 kV, the electrical plasma power reaches a value as high as 128 W. For comparison, the Figure 4b reports the values of the PTJ electrical power considering the same 5 configurations. Here, whatever the values of $V_0$, none of the configurations induces a change in the power which remains as low as 21 W at 9kV. In all configurations and whatever the APPJ, these power values are obtained for $V_0=9$ kV while the plasma plume always touches the target.

### (a) Plasma gun

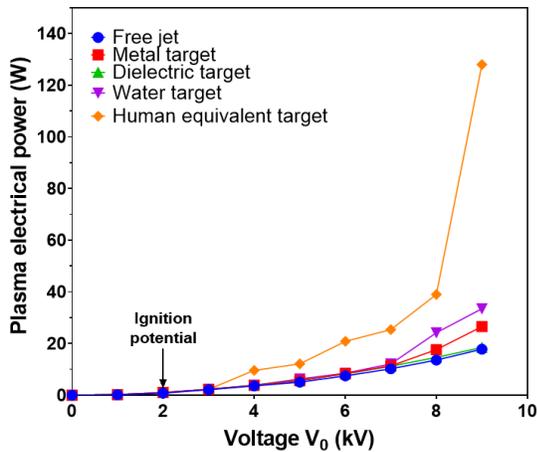

### (b) Plasma Tesla Jet

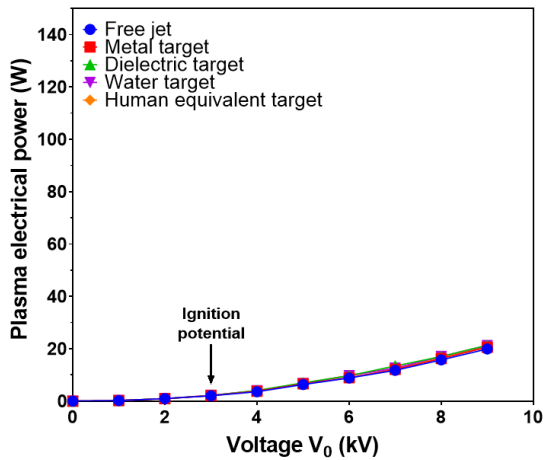

*Figure 4. Electrical plasma power per pulse as a function of the plasma voltage considering free jet or jet-target configurations for (a) Plasma gun and (b) Plasma Tesla Jet. Experimental conditions: helium flow rate=1000 sccm, frequency=30 kHz, duty cycle=14 % and gap=10 mm.*

To understand why plasma electrical powers are so different in PTJ and PG, the time profiles of the currents associated with each of these APPJs are analyzed: $I_{plume}$, $I_{gr}$ and $I_{gene}$ (Figure 3), keeping in mind that $I_{plume}$ is the current of interest owing to its interaction with the target. For the sake of clarity, we focus our analysis on two parameters characterizing pulse currents: (i) the maximum value of the instantaneous current for a pulse duration, (ii) the pulse current duration registered after application of the voltage pulse. These two parameters are measured for the two APPJ considering the 5 configurations, as shown in Figure 5.

With the PG, the maximum values of $I_{plume}$, $I_{gr}$ and $I_{gene}$ remain the same whatever the configuration. However, the pulse current duration is target-dependent, with for example $I_{plume}$ values as long as 1.3 µs and 1.8 µs for metal and water targets respectively and more than 4.0 µs using the EEHB target. On the contrary, with the PTJ, the maximum magnitude of current and the pulse current duration remain identical with/without targets and for electrical power values always equal or lower than with the PG. Considering the EEHB configuration at 9 kV introduced in Figure 4, the electrical plasma power of 128 W obtained with the PG results from pulse current durations longer than 4 µs while the power of 21 W obtained with the PTJ results from pulse current durations maintained as low as 0.9 µs. Owing to its ability to maintain a temporally narrow and stable pulse current over time, the PTJ provides a plasma electrical power which is not time or target dependent.

In this study, only the electrical power deposited in the grounded EEHB target is accurately measurable, as shown in Figure 6. In free jet and dielectric target configurations, $I_{target}=0$ for the reasons indicated in section 2.2. For the floating targets (metal, water), the figure 6 does not include $I_{target}$ since an electrical power can still locally dissipate by Joule effect although not measurable. For voltage values close to the ignition potential ($V_{ignit}=2$ kV for the PG and $V_{ignit}=3$ kV for the PTJ), the plume is constituted by a visible region (whose emissivity gradually vanishes as one moves away from the capillary), immediately followed by the optically transparent post-discharge region, as sketched in Figure 2. Assuming that the target is "touched" by the post-discharge and not by the plasma plume (see Figure 2), it turns out that a current can still be collected by the target, whatever its configuration. Two mechanisms could explain this target polarization : (i) between the plasma plume and the target, the post-discharge can be modelled as a virtual capacitor where – for example – the electrostatic field accumulates a certain amount of positive charges on the target while an equal amount of negative charges is accumulated on the plasma plume, (ii) the front of the ionization wave may be too poorly emissive to be detected by OES. As a result, in some cases, one might expect the target to be in contact with the post-discharge while it is actually in contact with the undetectable part of the plasma plume.

Increasing $V_0$ allows to extend the plume until a critical value, $V_{bridge}$, where the plume creates an optical contact with the target, i.e. the post-discharge totally vanishes. As illustrated in Figure 6, the electrical power deposited in the target increases from 1 mW to 70 mW for the PG-EEHB configuration (between $V_0=V_{ignit}=2$ kV and $V_0=5$ kV) and from 5 mW to 24 mW for the PTJ-EEHB configuration (between $V_0=V_{ignit}=3$ kV and $V_0=6$ kV). Then, once the optical bridging is obtained, the electrical power still deposits into the target although in a very different way: with the plasma gun, $P_{target}$ increases from 160 mW to 62 W (between $V_{bridge}=6$ kV and $V_0=9.0$ kV) while for the plasma Tesla jet $P_{target}$ remains close to 45 mW (between $V_{bridge}=7$ kV and $V_0=9.0$ kV). In that respect, the PTJ appears as a safe device for *in vivo* campaigns: it can be supplied with larger voltage without inducing any electrical hazard, offering the additional advantage of a more affluent gaseous chemistry as detailed in the next section.

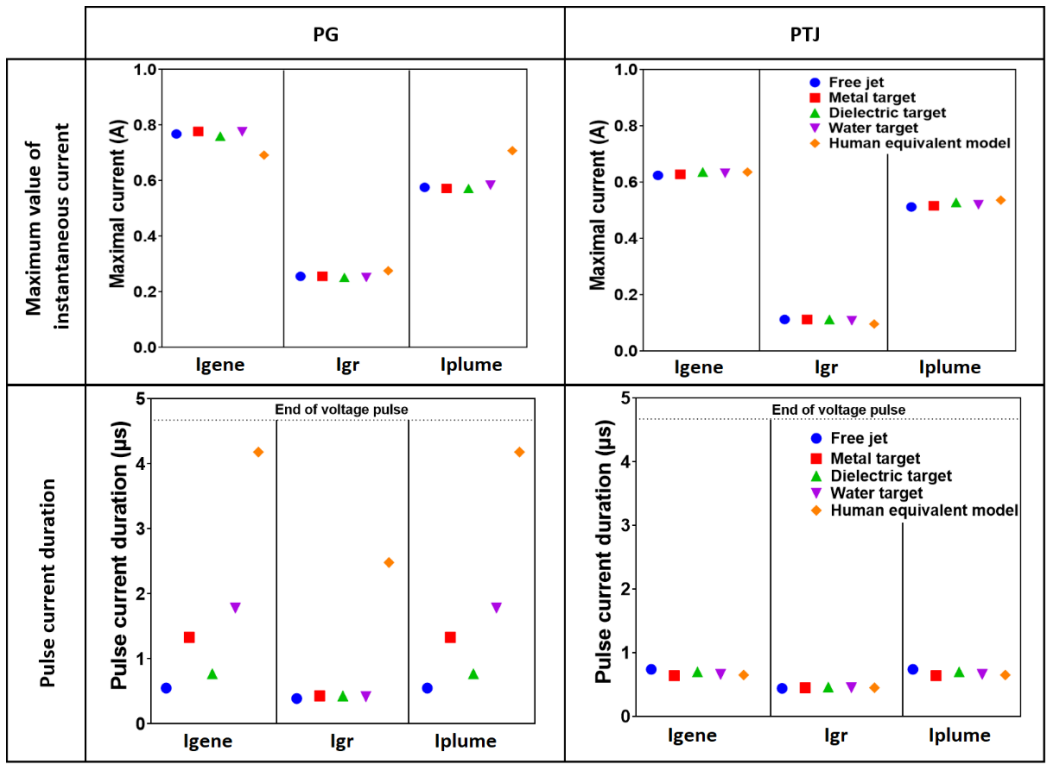

*Figure 5. Maximum values of current and pulse current durations for Plasma Gun and Plasma Tesla Jet with/without target interaction. Experimental conditions: $V_0$=9 kV, helium flow rate=1000 sccm, repetition frequency=30 kHz, duty cycle=14 %, gap=10 mm. The error bars are not visible due to the high reproducibility of the measurements (<2%).*

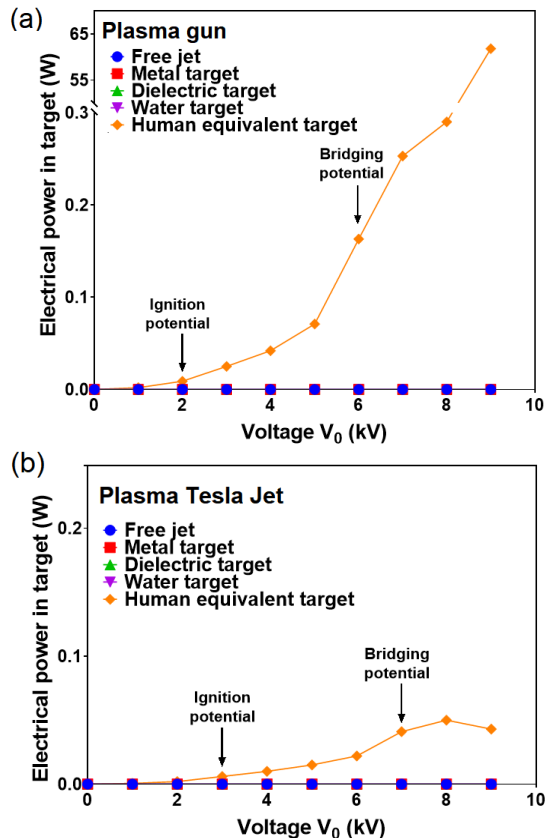

*Figure 6. Comparison of electrical power deposited in targets using (a) Plasma gun and (b) Plasma Tesla Jet. Experimental conditions: helium flow rate=1000 sccm, repetition frequency=30 kHz, duty cycle=14 % and gap=10 mm.*

## III.B. Interaction of plasma with targets: plasma parameters, short and long lifetime reactive species

Gas flowing dynamics of an APPJ is governed by parameters specific to the device itself (e.g. tube length, inner diameter of the tube, geometry), the plasma (e.g. temperature, gas flow rate, gas mixture) and the target (size, material, etc.). Such plasma device-target interaction has already been studied using Schlieren imaging in free jet configuration (Sarron et al 2013) as well as with conductive and dielectric targets (Darny et al 2017) (Li et al 2017) (Boselli et al 2014), providing qualitative and sometimes quantitative information on fluid dynamics using Toepler lenses (Traldi et al 2018). In this article, the flowing properties of the plume with/without target are investigated using space resolved mass spectrometry. Although this technique is slightly intrusive, it can be considered as complementary to Schlieren imaging while providing chemical information.

Two-dimension profiles of the plasma plume with/without target interaction are reported in Figure 7 for the PG and PTJ devices. The MS capillary is positioned side-on with respect to the jet axis. The two APPJ are supplied with a power of 12 W ($V_0$=7 kV, repetition frequency=30 kHz and duty cycle=14 %). Profiles are plotted in (r, z) coordinates with a spatial resolution of 1 mm. As sketched in the inset of Figure 7, the radial position r varies between 0 and 11 mm in all cases while the axial position z is probed on the 0-10 mm range in free jet and on the 0-8 mm range in presence of target.

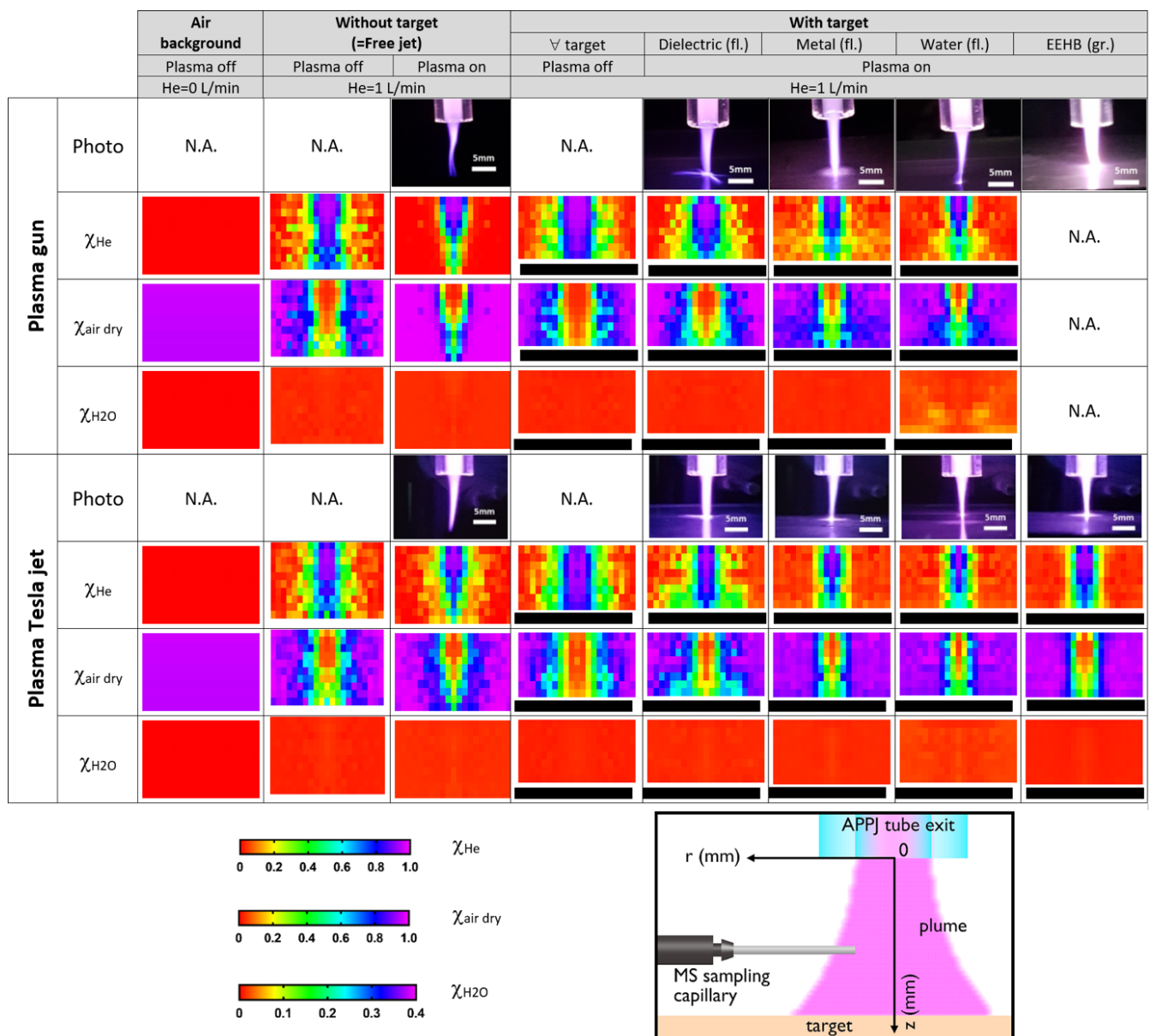

*Figure 7. 2D profiles of gas molar fractions for PG and PTJ with/without target interaction. Experimental conditions: Voltage 7 kV, helium flow rate=1000 sccm, frequency=30 kHz, duty cycle=14 % (electrical plasma power=12W) and gap=10 mm. The inset indicates how the MS capillary is positioned for perform the mapping.*

These profiles correspond to molar fractions of helium, air and water vapor, according to calculations reminded in the section *Experimental setup and methods*. The fluid dynamics properties are target-dependent:

- In the free jet configuration, PG and PTJ show similar $\chi_{He}$ and $\chi_{dry\_air}$ profiles depending on the on/off status of the plasma. When helium gas flows into the tube while the plasma is off, the plume shows a cylindrical profile with a diameter close to the inner tube diameter (4 mm). Then, igniting the plasma makes the plume becomes thinner as one moves away from the tube exit, giving rise to a cone-shape distribution.

- If a target interacts with a non-ionized gas flow of helium, whatever the target type and/or APPJ, the distribution of the plume remains cylindrical, with a diameter larger than the one obtained in free jet. Then, ionizing the gas into plasma enables the flow dynamics profile to be modified by the type of target and APPJ. In the case of the PTJ/dielectric target interaction, the diameter of $\chi_{He}$ is almost three times larger on the target than the inner diameter of the quartz tube (12 mm vs 4 mm, respectively). For the PTJ/metal target (grounded or floating), the helium flow propagates along a narrow cylinder channel, with a diameter of 4 mm, i.e. same as the tube inner diameter. The PG shows similar results while interacting with the metal target but in the case of the water target, the helium gas flow distribution is disrupted by vapor issued from the target heating. For the PG/EEHB interaction, no measurements have been performed since electrical power is too high and could damage the HV power supply.

Contrarily to what has been obtained with $\chi_{He}$ or $\chi_{dry\_air}$, the molar fraction of water vapor remains low ($\chi_{H2O}<0.03$) and does not show any preferential space profile whether in or out of the plasma plume. This random distribution of $\chi_{H2O}$ is obtained whatever the APPJ and in all target configurations (except with the aqueous target where higher $\chi_{H2O}$ may appear 2 mm above the interface).

The characterization of the plasma phase by mass spectrometry is completed by optical emission spectroscopy to identify radiative species of interest and calculate gas temperature. Whatever the APPJs with/without targets, same optical emission spectra are obtained, as indicated in Figure 8. Each spectrum is composed of the first negative and second positive systems of nitrogen ($N_2$), a band of hydroxide (OH) and lines of helium (He), oxygen (O) and hydrogen (H). Although no UV radiation of nitric oxide (NO) is detected by OES, nitrogen dioxide ($NO_2$) is evidenced by mass spectrometry.

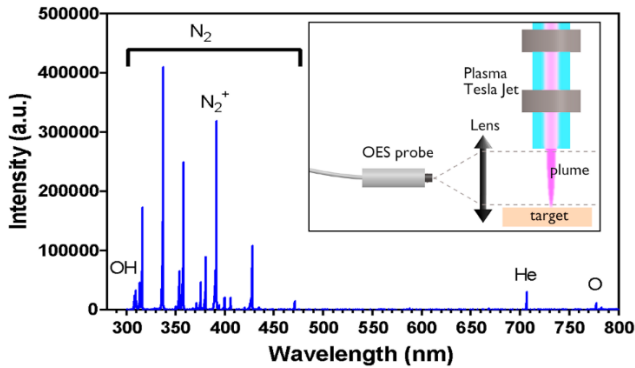

*Figure 8. Optical emission spectra of the plumes generated by plasma gun and plasma Tesla jet. The two spectra overlap whatever the APPJ with/without target interaction. The inset indicates how plasma emission is collected: along the jet axis (i.e. side-on) without spatial resolution*

To highlight the influence of targets on radiative species production/consumption mechanisms, all intensities are normalized with respect to the helium line at 706 nm. The Figure 9 shows the normalized intensities of the aforementioned species. With the PG, all radiative species ($N_2$ 337 nm, $N_2^+$ 391 nm, OH 309 nm, O 777 nm, He 587 nm and H 656 nm) are less intense than with the PTJ device except for the hydrogen radiation in free jet and dielectric target configurations. Moreover, the normalized intensity is more sensitive to the target configuration with the PG than with the PTJ. For example, nitrogen bands and oxygen line are not target-affected when PTJ is used. In both APPJ, OH intensity is higher with conductive target (i.e. metal and EEHB targets) than with water target, dielectric target and free jet. Normalized intensity of He indicates that both APPJs provide He 587 nm at a same level. Consequently, electron temperature is close in both APPJs except when the PG interacts with the EEHB target or with PTJ and water target. In these configurations, electron temperature is higher than in other configurations. Nevertheless, this electron temperature cannot be accurately assessed since the helium lines at 706 and 587 nm are too poorly emissive.

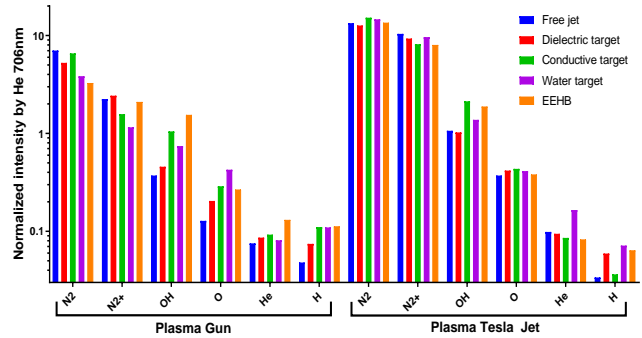

*Figure 9. Normalized intensity of OES lines and bands using PG and PTJ with/without target. $V_0=7$ kV, i.e. $P_{plasma}=12$ W for all configurations except for the PG-EEHB configuration where $P_{plasma}=26$ W).*

Rotational and vibrational temperature can be deduced from Boltzmann plot of OH(A-X) and $N_2$ (371, 375.5, 380.5 nm) respectively (Bruggeman et al 2014) (Ravari et al 2017). As reported in Table 1, the values of the plasma temperature indicate the generation of a non-equilibrium plasma with a vibrational temperature much higher than rotational temperature. Except for the PG/EEHB interaction, no significant difference can be observed between the two APPJs and whatever the target configuration. $T_{rot}$ is close to 310 K, i.e. close to the ambient temperature (293 K) while $T_{vib}$ is approximately 2500 K. During the PG/EEHB interaction, electrical power is two times higher than in other configurations. In the same case, rotational temperature turns around 1300 K and vibrational temperature 5500 K.

| | APPJ | Free jet | Targets | | | |
|---|---|---|---|---|---|---|
| | | | Dielectric | Conductive | Water | EEHB |
| $T_{rot}$ (K) | PG | 306 | 300 | 310 | 322 | 1277 |
| | PTJ | 310 | 311 | 304 | 304 | 300 |
| $T_{vib}$ (K) | PG | 2390 | 2446 | 2639 | 2747 | 5453 |
| | PTJ | 2545 | 2550 | 2659 | 2454 | 2606 |

*Tab.1: Rotational and vibrational temperature of plasma. Plasma power of the PG and PTJ is 12 W (voltage 7 kV) in all configurations except for the PG with EEHB target where it is 26 W.*

## III.C. Dermal toxicity survey: effects of PG on mice skin

Before assessing any plasma-induced antitumor effects, we have defined an experimental operating window inside which the values of the relevant experimental parameters can be changed without inducing deleterious effects on mice skin. The dermal toxicity assay has been achieved on the skin of immunocompromised mice exposed to PG or PTJ using values of duty cycle ($D_{cycle}$) at 14 % or 24 %, and a repetition frequency at 30 kHz and a gap of 10 mm (±2 mm). The uncertainty on this latter distance results from the mouse breathing.

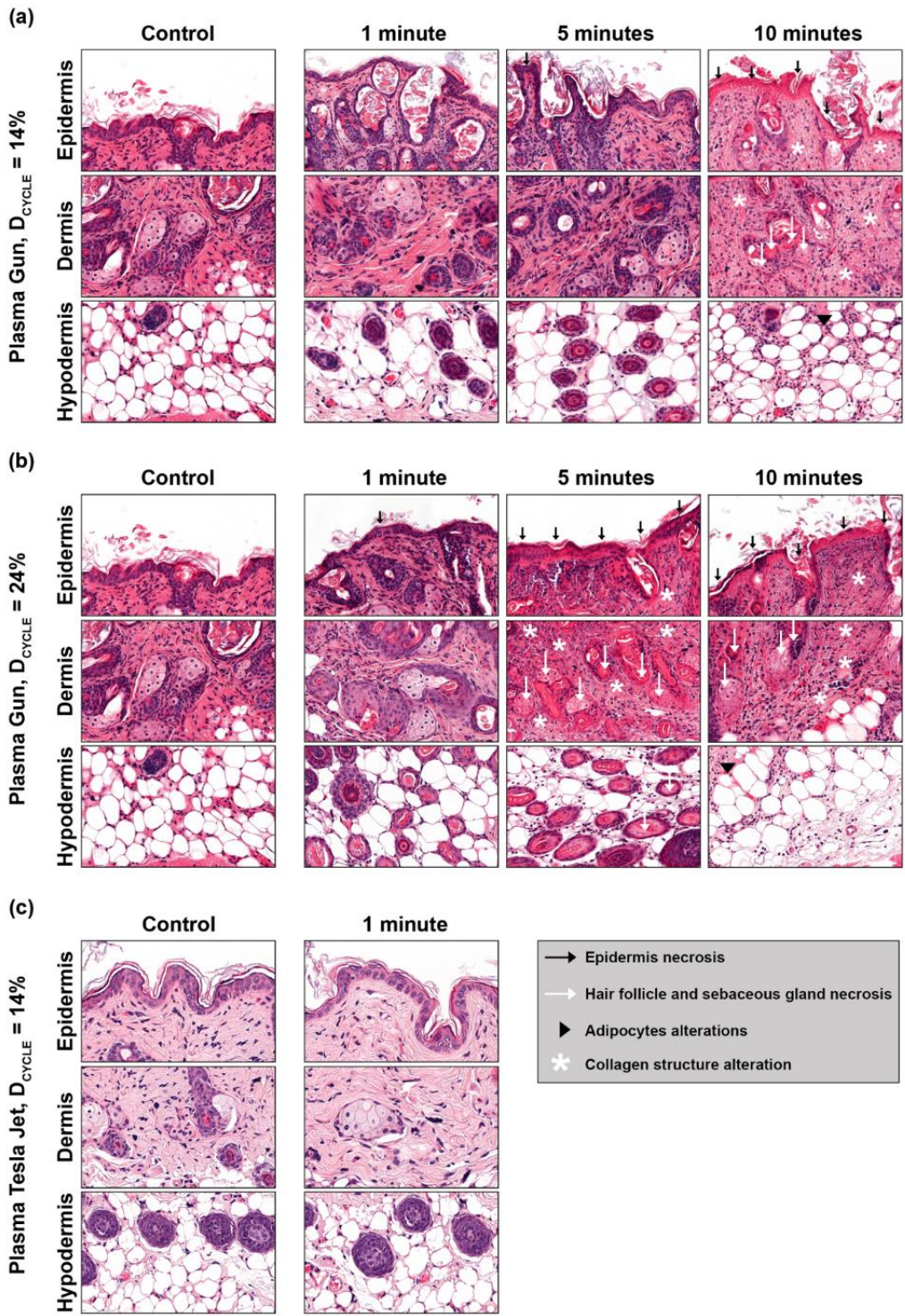

*Figure 10. Dermal toxicity test performed on the skin of immunodeficient mice. Representative histology (HE staining) of the skin after exposure to cold atmospheric plasma generated with (a) Plasma Gun, $D_{CYCLE}$=14 %, (b) Plasma Gun, $D_{CYCLE}$=24 % and (c) Plasma Tesla Jet, $D_{CYCLE}$=14 %. In all cases, $V_0$=7 kV. Magnification x40. n=4.*

In the case of the PG treatment, macroscopic analysis of the skin shows no major alteration of the skin in treated mice compared to the control group (data not shown). However, hematoxylin-eosin (HE) analysis of samples reveals skin alterations that are correlated with exposure time and duty cycle. As shown in Figure 10a, when used with a $D_{cycle}$=14 %, the PG does not induce any skin alterations after 1 min of treatment. However, after 5 minutes, HE staining reveals that half of the treated mice present slight alterations affecting mainly the epidermis with a prenecrotic aspect of keratinocytes in very focal areas, and very rarely alterations in the collagen structure of the superficial dermis. After 10 minutes of exposure, all treated mice show similar significant tissue modifications affecting all layers of the skin, with extensive necrosis of the epidermis, alteration in the collagen structure of the dermis, necrosis of hair follicles and sebaceous glands. Moreover, cellular modifications in the hypodermis are evidenced for one third of the mice exposed to PG during 10 min.

As shown in Figure 10b for $D_{cycle}$=24 %, exposing mice to PG during 1 min leads to skin alterations in half of them, i.e. focal prenecrotic lesions affecting the epidermis are observed. After 5 and 10 minutes of plasma exposure, HE staining indicates significant tissue necrosis affecting all layers of skin comparable to what is observed at the condition {$D_{cycle}$=14 %, time=10 min}. In very few cases, the microscopic alterations of the skin are observed macroscopically as a light skin redness that resolves within 24 h post-treatment, when the samples are collected.

Since PG treatment can induce skin alterations correlated with duty cycle and exposure time, the PTJ was evaluated only in the unharmful conditions defined with the PG : {$D_{cycle}$=14 %, time=1 min} (Figure 10c). The PTJ does not induce macroscopic and microscopic skin changes. Indeed, no epidermis, dermis or hypodermis alterations are evidenced histologically, all skin layers being strictly comparable with controls. Of note, only one mouse displays in a single focal area some very slight nuclei and cytoplasmic alterations involving keratinocytes of the epidermis (data not shown).

## III.D. Antitumor effects of PG and PTJ treatments on mice cholangiocarcinoma

Once the safety of the two plasma sources is clearly defined, their antitumor effects are evaluated in a subcutaneous xenograft tumor model performed with EGI-1 cholangiocarcinoma cells. After a few days, cancer cell proliferation generates a tumor localized under the skin, mimicking human tumor (Vaquero et al 2018). When tumor volumes reach an approximate volume as high as 200 mm³, two plasma gun treatments are achieved at days 13 and 20 for an exposure time of 1 min and $D_{CYCLE}$=14 %. As illustrated in Figure 11, the tumor volumes remain the same vs time whatever the control and plasma groups. Then, after leaving a 14 days refractory period, the tumors are treated using the PTJ at days 34 and 41 for an exposure time of 1 min and $D_{CYCLE}$=14 %. In this case, Figure 11 shows a strong reduction of tumor growth in the plasma group compared with the control group. At day 48, tumor volumes remain limited to 1250 mm³ for mice belonging to the plasma group vs 1730 mm³ for mice belonging to the control group. Experiments have not been carried out further to respect ethical protocols limiting tumor volume to less than 2000 mm³ for the two groups.

Altogether, the dermal toxicity test and the xenograft tumors experiments show that the PTJ is as safe as the PG and displays *in vivo* antitumor properties. The PTJ properties are more interesting since PTJ effect appears the day after PTJ treatment, i.e. at day 35 and that this effect is obtained at a very advanced stage of tumor development.

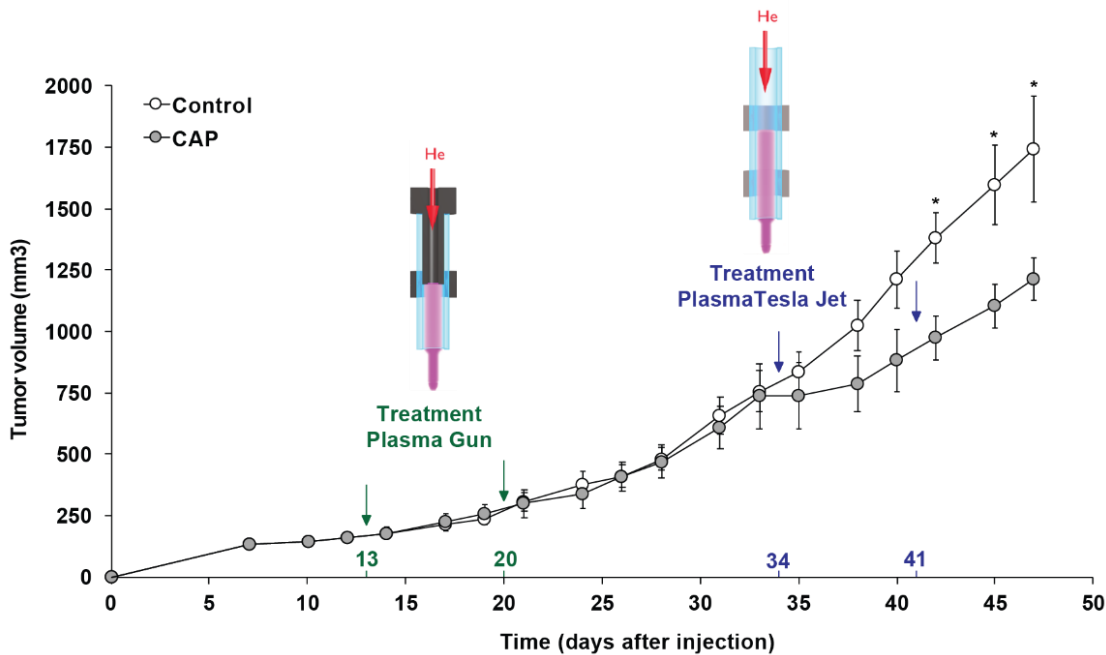

*Figure 11. Tumor volume of mice bearing EGI-1 cells treated with vehicle (white circles) or cold atmospheric plasma (grey circles) generated with Plasma Gun (arrows at days 13 and 20) and with Plasma Tesla Jet (arrows at days 34 and 41). Values are expressed as means ± SEM. *, p < 0.05; comparing CAP with vehicle. CAP, cold atmospheric plasma (n=4).*

# IV. Discussion

## IV.1. How targets influence plasma properties

Measuring the electrical current upstream (between HV generator and exciting electrode) and downstream (on grounded electrode) of the APPJ enables a deeper understanding of the plasma discharge mechanisms as well as a more accurate determination of electrical plasma power. The data presented in the "Results" section show that targets can greatly influence APPJ in different manners:

Targets exposed to a plasma gun can modify the propagation properties of the plasma, in particular the profile of the pulsed atmospheric plasma streams (PAPS). The PAPS exhibit a strong decay of their tail when the PG operates in free jet or in interaction with a dielectric target, while a longer one is obtained with ungrounded conductive targets. Transient thermal arcs can occur with grounded conductive target.

Targets can change fluid-dynamics properties of the plume. If the target is a dielectric (with a necessarily floating potential), the ionization wave propagates as short PAPS spreading over its surface. On the contrary, if the target is conductive, the plume propagation is performed along a cylinder-like channel, becoming even narrower as one gets closer to the target. In the case of a liquid target, plasma heating can convert part of the liquid into vapor, hence modifying the physico-chemical properties of the surrounding atmosphere.

The polarization status of the target impacts the propagation distance of the plume as well as the bridging potential, i.e. value of $V_0$ for which the plume is in optical contact with the target. The maximum propagation distance is as high as 20 mm if the target is grounded while only 12 mm if it is at floating potential. Besides, $V_{bridge}$ is always lower if the target is grounded rather than floating. Hence, in the case of the PG-metal target interaction, $V_{bridge}$=6 kV at floating potential vs 5 kV at grounded potential. Using the PTJ, the bridging potentials are even 1kV lower.

Chemical species production/consumption mechanisms are target-dependent, as shown in Figure 9 with reactive oxygen species, i.e. OH, O and H radicals.

Target configuration can strongly influence plasma parameters like electron density and plasma temperature (Klarenaar and al 2018), especially in the PG/grounded target configuration.

## IV.2. Plasma propagation mechanisms

With the PTJ, a dielectric barrier always separates HV electrode and plasma, hence preventing any transition to electric arc, whatever the target utilized.

With the PG, three target-dependent plasma propagation mechanisms can be identified in Figure 12:

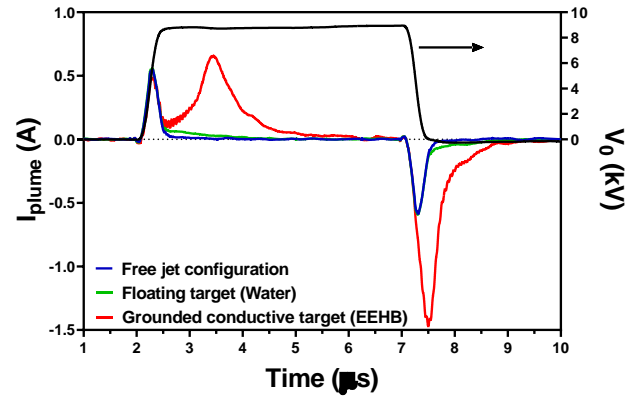

**Figure 12. Influence of targets on instantaneous current in PG plume. Experimental conditions: Voltage 9 kV, helium flow 1000 sccm, frequency 30 kHz and duty cycle 14 %. Voltage waveform is indicated in solid black line.**

Free jet configuration: the plasma behaves as an ionization wave characterized by an intense ionization front (i.e. local electric field of high magnitude) while its tail can be reasonably considered as negligible. Indeed, no ionization channel is detectable by electrical or imaging techniques once the wave front has propagated along the interelectrode distance. Such pulsed plasmas are referred as plasma bullets and generate current peaks that can be easily detected on the oscilloscope. In free jet, one peak of current appears at every edge (positive/negative) of the applied voltage, to decay a few µs later. These current peaks are slightly dissymmetric. Since the dielectric current corresponds to the product of the capacitance by the derivative of the applied voltage, its left wing is superimposed with the rise in voltage. If the voltage slew rate is low, dielectric and discharge currents can be time dissociated and so could be their peaks. If the voltage slew rate is high as in Figure 12, then they appear as a dissymmetric peak.

Floating targets (dielectric target, water or metal plate at floating potential): the plasma behaves like a Pulsed Atmospheric-pressure Plasma Stream (PAPS) (Robert et al 2015). Contrarily to the previous case, the wave front remains connected to the inner HV electrode through an ionization channel resulting from the propagation of the ionization wave and that can extend beyond the glass tube length, i.e. in ambient air. As a result, the current peak associated to a PAPS is highly dissymmetric: its right wing can decay on several µs (versus less than 0.5 µs on its left wing) and is the direct consequence of the PAPS tail. Once the ionization channel is entirely open by the wave front, the residual electrical charges bridge the HV electrode to the target. Such bridging lasts as long as the voltage between HV electrode and the floating conductive target becomes too low to keep open the plasma channel.

Grounded conductive target (e.g. human equivalent electrical circuit): 3 peaks appear at 2.3, 3.7 and 7.5 µs in Figure 12. The instantaneous current shows the profile of a RC series circuit. The instantaneous current is limited due to $C_{m1}$ (capacitor of measurement) and to the low duty cycle. Respecting these two conditions is a mandatory to prevent any cold discharge-to-arc transition.

In the case of the Plasma Gun, the conductive channel (resulting from the ionization wave propagation) can bridge the grounded target to the high voltage electrode. If the applied voltage is high enough, this bridging can lead to a discharge-to-arc transition. Then, the electrical power dissipated into the target can be limited if the plasma impedance is much higher than the target impedance. This condition can be satisfied either by reducing the inner diameter of the tube or by increasing the HV electrode-to-target distance.

To mimic the influence of this latter parameter on the power deposited on living models, the grounded EEHB target has been exposed to PG and PTJ for gaps distances ranging between 1 and 20 mm. The corresponding powers are reported in Figure 13. With the PG, the deposited electrical power strongly depends on the gap: for gaps lower than 11mm, power is as high as 480 W for $V_3$=2.0 kV and the plasma appears as an electric arc (see inset of Figure 13). Then, for gap longer than 11 mm, the dissipated electrical power is drastically reduced, ranging between 0.66 W at 12mm to 0.02 W at 20 mm. For higher gaps, plasma is no more bridged to the EEHB target. The power values given in Fig. 13 are much higher than those introduced in Fig. 4 and Fig. 6 since in this electrical setup, measurements have been carried out without $C_{m1}$ capacitor.

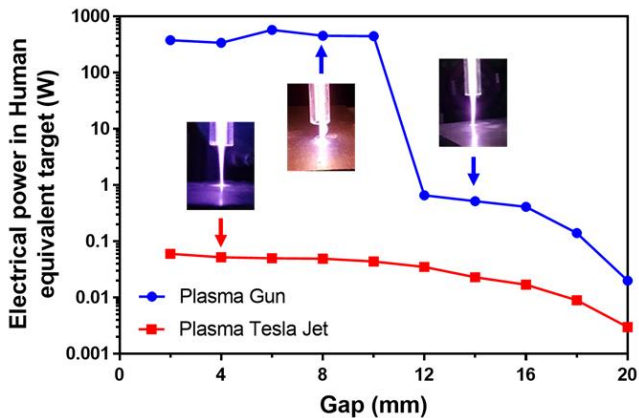

*Figure 13: EEHB target power as a function of the gap for plasma gun and plasma Tesla jet devices. Experimental conditions: $V_0$=7 kV, helium flow rate=1000 sccm, repetition frequency=30 kHz and duty cycle=14 %. No measurement capacitor $C_{m1}$ is present in the electrical circuit.*

As a result, the current cannot be estimated in the arc even if one can estimate its effective value in the EEHB target using $I_{target,eff} = \sqrt{\frac{1}{T} \cdot \int_T \frac{V_{target}^2(t)}{R_{target}} \cdot dt}$. Hence, with the PG operating at lower gaps, $I_{target,eff}$=500 mA (P=800W) while for higher gaps it is only 10 mA (P=0.6 W).

The electrical power deposited in the target is always lower with PTJ than with PG, since it ranges from 60 mW (2 mm) to 3 mW (20 mm). Moreover, the $V_3$ potential (and therefore the voltage along the EEHB target) is less than 1.5 V.

## IV.3. Comparing PG & PTJ in regard of their physical properties

From an energy balance point of view, the electrical power supplied by the HV generator is the sum of thermal power, chemical power and radiative power consumed by the plasma. If the precise quantification of each contribution is beyond the scope of the present article, one can however suggest lines of thoughts for future work. First, it would be important to verify whether the chemical contribution – and more specifically the contribution of the short lifetime reactive species – is richer and more selective with PTJ than with PG, therefore contributing into higher antitumor effects. Since the emission band and lines of $N_2^+$, OH and O in Figure 9 are higher with PTJ than with PG, one may assume a richer PTJ-induced radical chemistry. However, such statement remains an assumption since quenching processes must also be assessed using a collision radiative model to calculate atomic state distribution functions versus particle densities and temperatures (Hartgers et al 2001). Mass spectrometry measurements have been performed in this work but its 1m long flexible capillary make this technique not accurate enough to conclude on transient chemistry. Second, one can reasonably consider the radiative power much higher with PG than with PTJ if one refers to the absolute optical emission of the He line at 706 nm. Indeed, its intensity ranges between 75000-93000 a.u. with PG vs 29000-41000 a.u. with PTJ. Third, thermal power is similar in both APPJ with/without target except the EEHB target, as reported in Table 1. To summarize, one may assume that PG enables higher radiative transfers while the PTJ better promote chemical consumption/production mechanisms. Such dissimilar behaviors remain hypothetical and could result from two different electron energy distribution functions, one specific to the PG and the other to the PTJ.

## IV.4. Comparing PG & PTJ in regard of safety issues

Connecting a conductive target to the ground allows the generation of a dynamic electric field between the target surface and the ionization wave's front. As a result, the bridging potential is obtained at lower value, whatever the APPJ. With the PTJ, such bridging is inconsequential in terms of electrical hazards on *in vivo* models since the plume propagates as short/long PAPS. However, it becomes critical if one utilizes a plasma gun configuration without precautions. Indeed, as soon as $V_0$ reaches the $V_{bridge}$ value, an electric arc is likely to appear since no dielectric barrier separates the HV electrode to the grounded target. Then the gas temperature increases to values far larger than the threshold authorized for our *in vivo* experiments (313 K, 40 °C). To reduce arcing, one can limit the current from 500 mA to 10 mA by placing a capacitor like $C_{m1}$ downstream of the APPJ, as shown in Figure 3. In our case, the gas temperature is lowered although its value (approx. 1000 K) remains too elevated for medicine applications. To prevent the arcing without modifying the PG electrode configuration, the best option is to increase the plasma impedance by elongating the post-electrode length; for example, using a longer dielectric tube as successfully demonstrated upon *in vivo* campaigns (Brullé et al 2012).

Finally, gas dynamics profiles achieved in the post-electrode region are very different depending on the non-ionized or ionized status of the carrier gas. If this behavior remains poorly understood, several mechanisms are foreboded including gas heating, local pressure increase, gas transport properties or momentum transfer between ions and neutrals (Boselli et al 2014).

## IV.5. Comparing PG & PTJ antitumor properties

To understand why one of the two plasma sources validates an anti-tumor effect and not the other, we propose to briefly compare them at the light of the usual cold plasma properties, i.e. radiative, thermal, chemical, electrical and gas flow properties.

According to Table 1, the gas temperature (neutral species) is always close to 305 K ( $\pm 6$ K) whatever the plasma source (with the exception of the arc regime which cannot be applied on living organisms). Therefore, we assume that for such a low value, the temperature cannot induce antitumor effect owing to the absence of such results with the PG. Regarding flow properties, Figure 7 shows profiles that significantly differ when the target is changed but not when the PG is replaced with the PTJ. Again, in our own experimental conditions, the impact of flow properties can be considered as negligible to induce antitumor effects.

The balance between radiative and chemical properties could partly explain the anti-tumor effects demonstrated in Figure 11. As previously noted, the radiative properties are stronger with PG than with PTJ, whereas more chemical species may be produced with PTJ, likely to explain the tumor size reduction. In our *in vivo* experiments, the tumors were ectopically grafted on mice, i.e. covered by a thin skin layer likely to mitigate the diffusion of exogenous radicals from plasma. Since reactive species can be delivered several millimeters into tissues (Szili et al 2018), it is important to identify if those detected with PTJ and PG are likely to induce antitumor effects:

The hydroxyl (OH) radical is the most electrophilic ROS with high reactivity. It can cause oxidative damage to DNA, proteins and lipids as long as it is produced in their vicinity (Hadi et al 2010) (Cadet & Davies 2017). In our research works, even if OH radicals are significantly produced with the two APPJ, their therapeutic potential remains questionable in inducing antitumor effects.

Low concentrations of extracellular singlet oxygen can inactivate catalase on the membrane of tumor cells and thus abrogate the antioxidant activity of one of the central molecules of tumor cells (Riethmüller et al 2015). Although produced in low amounts with our APPJ, the role of O radical as antitumoral agent must not be underestimated. In the vicinity of inactivated catalase, it could prevent NO from oxidation and prevent $H_2O_2$ and peroxynitrite (constantly produced outside of tumor cells) to be decomposed (Bauer 2016). Then, the subsequent protonation of peroxynitrite into peroxynitrous acid can enable the production of intracellular $NO_2^\bullet$ and hydroxyl radicals.

To the best of the authors knowledge, the H radical is not described in the literature as a candidate likely to induce strong anticancer effects. Besides its production into the plasma phase remains very low as shown in Figure 9. For these reasons, H radicals can reasonably consider as playing a negligible role in the antitumor effects highlighted in Figure 11.

As reminded by D. Graves (Graves 2014), nitric oxide (NO) is a biologically significant molecule that can induce several pivotal effects, e.g. immune modulation of tumor growth, modulation of angiogenesis and inhibition of cell respiration (Morbidelli et al 2019) (Janakiram & Rao 2015). Although NO has not been investigated in the preset study, further works could be carried out to generate it on purpose, as selectively as possible.

Finally, the plasma electrical properties are also assumed to play an important role in the antitumor effect. Depending on whether PG or PTJ is used on the same target, the resulting $V_{target}=f(t)$ profiles could highlight substantial differences (change in voltage polarity, change in pulses duration, …) that are still under investigation.

# V. Conclusion

Simple targets (dielectric plates, grounded/floating metal plates) are useful to quickly and cheaply determine the plasma properties of APPJ devices interacting with a biological system. However, the values of the measured plasma parameters have to be considered with great care. Since the impedance of these targets is different from the impedance of any living model, the measured plasma properties can only be used to benchmark the APPJ among themselves, without possibly predicting their accurate values if a living model is treated instead. In particular, $P_{plasma}$ and $P_{target}$ values can be unintentionally underestimated, as demonstrated in this work with the PG, since the maximum value of instantaneous current as well as the pulse current duration depend on the target type. These discrepancies are even stronger if one considers a target mimicking the electrical properties of the human body. Indeed, electrical hazards are only detectable with the EEHB target and not with simple dielectric/metal targets. The reason is the propagation of the PAPS which is target-dependent.

On the contrary, electrical properties of the PTJ are not affected by the target itself owing to the dielectric glass tube separating the plasma bulk from the HV electrode and which always plays the role of dielectric barrier. Thus, strong stability and reproducibility are guaranteed, PAPS propagation does not depend on the target type and no electrical hazards can occur like transition to arc. Another advantage of the PTJ is its ability to produce higher amounts of reactive species compared with a PG operating in the same conditions. These strengths must however be nuanced with regard to the enhanced radiative properties of the PG.

The dermal toxicity survey has shown that potential deleterious effects can be obtained on the skin for long exposure times (>5 min) and high duty cycles: prenecrotic aspect of keratinocytes in very focal areas of the epidermis followed by necrosis of the epidermis, alteration in the collagen structure of the dermis, necrosis of hair follicles and sebaceous glands. To prevent such effects, optimal values have been set as follows: duty cycle=14 %, repetition frequency=30 kHz, magnitude=17kV, gap=10 mm and exposure time=1 min.

In conclusion, we have engineered a cold atmospheric plasma device showing a therapeutic efficiency for the treatment of cholangiocarcinoma. If the plasma gun has already been successfully applied on murine models to induce antitumor effects, the PTJ appears today as a promising alternative for the treatment of specific/aggressive cancers as cholangiocarcinoma. Further investigation is ongoing to dissect the cellular and molecular mechanisms involved in the antitumoral effect. Altogether, the study should improve the usefulness of the device and be able to provide new anticancer treatment opportunities to patients suffering from cholangiocarcinoma as well as other aggressive cancer.

# VI. Acknowledgements

This work has been done within the LABEX Plas@par project and received financial state aid managed by the Agence Nationale de la Recherche, as part of the programme "Investissements d'avenir" (ANR-11-IDEX-0004-02) and within the Emergence @ Sorbonne Universités 2016 fundings. Also, it has been supported by the « Région Ile-de-France " (Sesame, Ref. 16016309), Sorbonne Université Platform program and by French network GDR 2025 HAPPYBIO. The authors acknowledge Tatiana Ledent from Housing and experimental animal facility (HEAF) and Fatiha Merabtene from the histomorphology Platform, UMS 30 Lumic, Centre de recherche Saint-Antoine (CRSA).